\documentclass[pra,amssymb,amsmath,twocolumn,superscriptaddress]{revtex4}

\usepackage{epsfig,amsmath,amssymb,bbm}
\usepackage{graphicx}

\newcommand{\identity}{\mathbbm{1}}
\newcommand{\ket}[1]{\left| #1 \right\rangle}

\newcommand{\proj}[1]{| #1 \rangle \! \langle #1 |}
\DeclareMathOperator{\trace}{Tr}

\begin{document}

\title{A No-Go Theorem for Gaussian Quantum Error Correction}

\author{Julien Niset} 
\affiliation{QuIC, Ecole Polytechnique, CP 165, Universit\'e Libre de
Bruxelles, 1050 Brussels, Belgium}

\author{Jarom\'{\i}r Fiur\'{a}\v{s}ek} 
\affiliation{Department of Optics, Palack\'{y} University, 17. listopadu 50, 77200 Olomouc, Czech Republic}

\author{Nicolas J. Cerf} 
\affiliation{QuIC, Ecole Polytechnique, CP 165, Universit\'e Libre de
Bruxelles, 1050 Brussels, Belgium}
\affiliation{Research Laboratory of Electronics, Massachusetts Institute of Technology, Cambridge, MA 02139}

\begin{abstract}
It is proven that Gaussian operations are of no use for protecting Gaussian states against Gaussian errors in quantum communication protocols. Specifically, we introduce a new quantity characterizing any single-mode Gaussian channel, called \textit{entanglement degradation}, and show that it cannot decrease via Gaussian encoding and decoding operations only. 
The strength of this no-go theorem is illustrated with some examples of Gaussian channels.
\end{abstract}


\maketitle

Quantum information processing based on continuous variables has attracted much attention over the recent years due to both its conceptual simplicity and experimental advantages. In particular, the set of Gaussian states and operations have been shown to enable many quantum information primitives, such as teleportation \cite{teleportation}, key distribution \cite{qkd}, and cloning \cite{cloning}. Interestingly, when the quadratures of the electromagnetic field are used to carry information, the entire set of Gaussian operations can be implemented by combining passive linear optical components such as beam splitters and phase shifters together with squeezers and homodyne detection followed by feedforward. All these elements are, up to some degree, readily accessible in today's optical laboratories. However, manipulating Gaussian states with Gaussian operations also leads to some limitations. Probably the most significant one is the impossibility to distill entanglement from Gaussian entangled states with Gaussian local operations and classical communication \cite{distil1,distil2,distil3}. As a result, some important quantum primitives, such as quantum repeaters, cannot be implemented within the Gaussian regime and hence require the use of experimentally more demanding non-Gaussian resources, such as photon subtraction \cite{ngaus_st} or de-gaussification operations \cite{ngaus_op}. Given the present state of technology, understanding what is possible or not within the Gaussian regime is thus of a great importance as it underpins the ``main stream'' use of optical continuous variables in quantum information protocols.

Recently, several schemes have been developed to fight noise and losses in continuous-variable quantum transmission lines. Given the well-known connection between quantum error correction and entanglement distillation for discrete-variable quantum systems \cite{bennett}, it was implicitly assumed that correcting Gaussian errors with Gaussian operations would be impossible. Logically, these schemes were thus all focused on non-Gaussian error models, such as discrete errors \cite{Braunstein98,Lloyd98}, phase-diffusion noise \cite{franzen}, probabilistic phase-space kicks \cite{heersink}, or probabilistic losses \cite{wittmann,niset}. However, the sole existence of a no-go theorem for Gaussian error correction had yet remained unresolved.

In this Letter, we address this problem by introducing a new intrinsic feature of single-mode Gaussian channels, which
we call \textit{entanglement degradation} $D$. This parameter characterizes the extent to which the channel degrades entanglement when acting on one half of 
two-mode squeezed vacuum at the limit of infinite squeezing. By exploiting a connection between quantum error correction and entanglement distillation in the Gaussian regime, we prove that $D$ can never decrease when one is restricted to Gaussian encoding and decoding operations. Our result is thus of the form of a \textit{no-go theorem}, establishing the impossibility of improving the transmission of Gaussian states in a Gaussian channel with Gaussian error correction only.

\begin{figure}[t]
  {\includegraphics[width=0.5\textwidth]{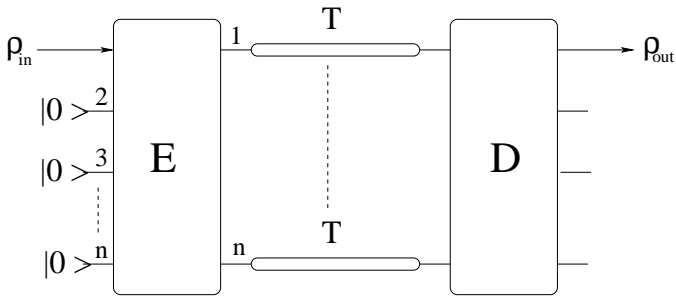}}
  \caption{Scheme of a Gaussian Error Correcting Code $G(n,E,D)$ for a Gaussian channel $T$, where $E$ and $D$ are $n$-mode unitary Gaussian operations used for encoding and decoding, respectively.}
\label{fig:gec}
\end{figure}

Let us briefly remind the Gaussian formalism. Any $n$-mode Gaussian state is completely characterized by its first and second moments $\mathbf{d}$ and $\gamma$, respectively. Introducing the vector of quadratures $\mathbf{r}=(x_1,p_1, \ldots, x_n,p_n)$, the coherent vector and covariance matrix read $d_{j}=\langle r_j\rangle$ and $\gamma_{ij}=\langle r_ir_j+r_jr_i\rangle-2d_{i}d_{j}$. A quantum Gaussian channel is a trace-preserving completely-positive map $T$ that transforms Gaussian states into Gaussian states according to $\rho\rightarrow T(\rho)$. It can be understood as resulting from a Gaussian unitary operation $U$ (associated with a quadratic bosonic Hamiltonian) acting on the state $\rho$ together with its environment in a Gaussian state $\rho_E$, i.e., $T(\rho)=\trace_E U(\rho\otimes\rho_E)U^\dagger$, where $\trace_E$ denotes partial trace with respect to the environment \cite{bookcerf}. Gaussian channels are known to model many physical lines, e.g., the transmission through a lossy optical fiber. At the level of covariance matrices, the action of $T$ is completely characterized by two matrices $M$ and $N$, 
\begin{equation}
 \gamma \rightarrow M \gamma M^T + N,
\end{equation}
where $M$ is real and $N\geq 0$ is real and symmetric. In the case of a single-mode channel, the condition of complete positivity of the map implies that
\begin{equation}\label{cpmap}
 \det N \geq (\det M - 1)^2,
\end{equation}
which means that any map $\gamma \rightarrow M \gamma M^T$ can be approximately realized provided that sufficient noise $N$ is added. An important subclass of Gaussian operations are symplectic transformations, corresponding to $N=0$ and $\det M=1$. Well-known examples of such operations are phase-shifters and squeezers, with symplectic matrices
\begin{equation}
 M_\mathrm{PS} = \left(
\begin{array}{cc}
\cos \theta & \sin \theta\\
-\sin \theta & \cos\theta
\end{array}
\right),
\hspace{5mm}
M_\mathrm{Sq}= 
\left(
\begin{array}{cc}
e^{-r} & 0\\
0 & e^r
\end{array}
\right)\, .
\end{equation}
Let us now define a Gaussian Error-Correcting Code (GECC) associated with the Gaussian channel $T$, as depicted in Fig. \ref{fig:gec}. It consists of a finite number $n-1$ of ancillas in a vacuum state $\ket{0}$, Gaussian unitary operations for encoding $E$ and decoding $D$ acting each on $n$ modes, and $n$ uses of the channel $T$. The code is denoted by $G(n,E,D)$, and its overall effect is to turn the Gaussian channel $T$ with matrices $M$ and $N$ into a Gaussian channel $T_\mathrm{GC}$ with matrices $M_\mathrm{GC}$ and $N_\mathrm{GC}$. 

We are now ready to turn to the proof of our no-go theorem, which can be greatly simplified by introducing two Lemmas. For discrete systems, it is well known that any error-correcting code is equivalent to a one-way entanglement distillation protocol, and vice-versa \cite{bennett}. For continuous variables, this relation is not as straightforward, but one can nevertheless prove the following lemma: 

\textbf{Lemma 1:} If $|\phi_{r}\rangle$ is a two-mode squeezed vacuum state with squeezing parameter $r$, the code $G(n,E,D)$ transforming the Gaussian channel $T$ into the Gaussian channel $T_{GC}$ is equivalent to a one-way 
protocol transforming $n$ copies of the state $\chi=\lim_{r\rightarrow \infty}\identity\otimes T(\proj{\phi_{r}})$ into one copy of the state $\rho_{\mathrm{out}}= \lim_{r\rightarrow \infty}\identity\otimes T_{GC}(\proj{\phi_{r}})$ by local Gaussian operations only. 

\textbf{Proof.} Our main tool is the well-known isomorphism between CP maps and positive operators \cite{isomo}. In particular, to any Gaussian CP map $T$ acting on a Hilbert space ${\cal H}$ corresponding to one mode, one can associate a Gaussian positive operator $\chi$ on ${\cal H}\otimes {\cal H}$ defined as
\begin{equation}
 \chi=\lim_{r\rightarrow \infty}\identity\otimes T(\proj{\phi_{r}}),
\end{equation}
where $|\phi_r\rangle=\sqrt{1-\tanh^2(r)}\sum_n \tanh^n(r)|n,n\rangle$ is a two-mode squeezed vacuum state. Acting with $T$ on a Gaussian state $\rho$ can thus be seen as teleporting $\rho$ through the quantum gate defined by the resource state $\chi$ \cite{distil3}. It follows that the $n$ uses of $T$ appearing in the GECC can be replaced by $n$ teleportations associated with the resource state $\chi$. Note that the operations involved in the teleportation, i.e., Bell measurement, one-way classical communication, and displacement all maintain the overall Gaussian character of the scheme. If the input of the GECC is now chosen to be one half of the state $|\phi_r\rangle$, then $G(n,E,D)$ is turned into a one-way Gaussian protocol that transforms $n$ copies of state $\chi$ into one copy of state $\rho_{\mathrm{out}}= \identity\otimes T_\mathrm{GC}(\proj{\phi_{r}})$. 
The protocol is the following: Alice prepares the entangled state $|\phi_r\rangle$ and $n-1$ ancillas, then applies the Gaussian operation $E$ on the ancillas and one half of $|\phi_r\rangle$. Next, she performs $n$ Bell measurements using the $n$ copies of the resource state $\chi$ and communicates the results to Bob. Bob displaces his shares of the $n$ resource states accordingly, and applies the Gaussian operation $D$. Alice and Bob now share one copy of the state $\rho_{\mathrm{out}}$. In particular, if Alice prepares the entangled state $\lim_{r\rightarrow \infty}|\phi_r\rangle$, the state they finally share is $\rho_{\mathrm{out}}=\lim_{r\rightarrow \infty} \identity\otimes T_\mathrm{GC}(\proj{\phi_{r}})$.
$\blacksquare$

\begin{figure}[t]
  {\includegraphics[width=0.5\textwidth]{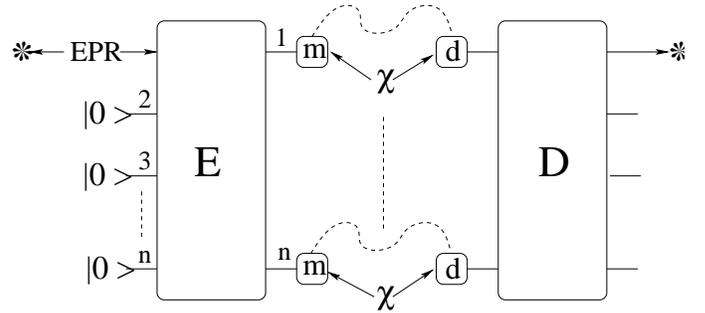}}
  \caption{From a GECC to Gaussian entanglement distillation. m: Bell measurement, d: displacement.}
\label{fig:ed}
\end{figure}

The preceding Lemma does not say anything about the entanglement of the resource state $\chi$ and final state $\rho_{\mathrm{out}}$, 
which is why we refered to a one-way protocol, not a one-way \textit{entanglement distillation} protocol. For the protocol to be a true entanglement distillation protocol, one has to show that it increases entanglement, i.e., that $E[\rho_{\mathrm{out}}]>E[\chi]$ for some entanglement measure $E$. This is addressed by the following Lemma.

\textbf{Lemma 2:} Given a Gaussian channel $T$ with matrices $M$ and $N$ acting on one half of the entangled state $\rho_{\mathrm{in}}=\lim_{r\rightarrow \infty}\proj{\phi_r}$, the entanglement of the output state 
$\rho_{\mathrm{out}}=\lim_{r\rightarrow \infty} \identity\otimes T(\proj{\phi_{r}})$
is completely characterized by the \textit{entanglement degradation} of the channel
\begin{equation}
 D[T] = \min\left\{\frac{\det N }{(1+\det M)^2},1\right\}.
\end{equation}
In particular, the logarithmic negativity is the decreasing function   $E_N[\rho_{\mathrm{out}}]=-\frac{1}{2}\log D[T]$. 

\textbf{Proof.} Let us first assume that $\det M > 0$. Without restriction, we can choose $M=\eta \identity$ since the channel can always be transformed into another Gaussian channel with $M'=SVMU$ and $N'=SVNV^TS$ by adding two phase shifts of symplectic matrices $U$ and $V$ at the input and output, respectively, followed by a single-mode squeezer of symplectic matrix $S$. Note that these operations are local, so they do not affect the entanglement properties of the channel. By singular value decomposition, $U$ and $V$ can be chosen such that $VMU$ is diagonal. Then, tuning the squeezing appropriately can make $M'$ proportional to the identity, i.e., $M'=\eta \identity$. Importantly, the determinant of symplectic matrices being equal to unity, $\det M' =\det M$ and $\det N'= \det N$. 

Let us now consider the action of $T$ on one half of the state $\proj{\phi_r}$
with covariance matrix $\gamma_{\mathrm{in}}^{(r)}$. Recalling that covariance matrices of two-mode Gaussian states can be decomposed in four $2\times2$ blocks, we easily find the input and output covariance matrices to be
\begin{equation}\label{tms}
 \gamma_{\mathrm{in}}^{(r)} = \left(
\begin{array}{cc}
A_r & C_r\\
C_r & A_r
\end{array}
\right),
\hspace{5mm}
\gamma^{(r)}_{\mathrm{out}}= 
\left(
\begin{array}{cc}
A_r & \eta C_r\\
\eta C_r & \eta^2 A_r + N
\end{array}
\right),
\end{equation}
with 
\begin{equation}
\nonumber A_r=\cosh(2r) \left(
\begin{array}{cc}
1 & 0\\
0 & 1
\end{array}
\right), 
\quad
C_r=\sinh(2r) 
\left(
\begin{array}{cc}
1 & 0\\
0 & -1
\end{array}
\right).
\end{equation}
Now, remember that the entanglement of a two-mode Gaussian state with covariance matrix 
\begin{equation}
\nonumber \gamma= 
\left(
\begin{array}{cc}
A & B\\
B^T & C \end{array}
\right)
\end{equation}
is fully characterized by the smallest symplectic eigenvalue $\nu_-$ of the partially transposed state \cite{adesso}. In particular, the logarithmic negativity is given by $E_N=\max\{0,-\log \nu_-\}$. One can calculate $\nu_-$ from $\gamma$ using
$$2\nu_-^2 = \tilde{\Delta} - \sqrt{\tilde{\Delta}^2-4 \det\gamma} ,$$
where $\tilde{\Delta}=\det A+\det C-2\det B$ \cite{ent}. For the output state $\rho^{(r)}_{\mathrm{out}}=\identity\otimes T(\proj{\phi_{r}})$ of covariance matrix $\gamma^{(r)}_{\mathrm{out}}$, a few lines of calculation yields
\begin{align}\label{sympl_inv}
\nonumber \tilde{\Delta}&= \cosh^2(2r)(1+\eta^2)^2 + O(\cosh(2r)),\\
\det \gamma^{(r)}_{\mathrm{out}}&=\cosh^2(2r)\det N +O(\cosh(2r)),
\end{align}
where we have used some known rules for the determinant of block matrices and the relation $\det(A+\lambda \identity)=\det A+\lambda \trace A + \lambda^2$, which is valid for $2\times2$ matrices. We can now calculate $\nu_{-}^2$ for the state $\rho_{\mathrm{out}}=\lim_{r\rightarrow \infty} \identity\otimes T(\proj{\phi_{r}})$. Using Eq. (\ref{sympl_inv}) and $\sqrt{1-x}= 1 - x/2+O(x^2)$, we find
\begin{align}
\nonumber \lim_{r\rightarrow \infty} 2 \nu_-^2&=\lim_{r\rightarrow \infty} \tilde{\Delta} - \sqrt{\tilde{\Delta}^2-4 \det \gamma^{(r)}_{\mathrm{out}}}\\
& =\lim_{r\rightarrow \infty}\frac{2\det \gamma^{(r)}_{\mathrm{out}}}{\tilde{\Delta}}
=\frac{2\det N}{(1+\eta^2)^2}.
\end{align}
Recalling that $\det M  = \eta^2$, we obtain the logarithmic negativity of the state $\rho_{\mathrm{out}}$
\begin{equation}
E_N[\rho_{\mathrm{out}}]=-\frac{1}{2}\log\left(\min\left\{\frac{\det N}{(1+\det M)^2},1\right\}\right)\, .
\end{equation}

Let us now consider a second class of channels, characterized by $\det M < 0$. An example of such channel is the approximate phase-conjugation map \cite{pc}. Using the same arguments as before, it is easy to show that we can restrict our attention to $M=\eta \varLambda$, where $\varLambda$ is a real diagonal matrix with $\det \varLambda = -1$.
The two symplectic invariants can again be easily calculated, and while $\det \gamma^{(r)}_{\mathrm{out}}$ is unchanged, now $\tilde{\Delta}= \cosh^2(2r)(1-\eta^2)^2 + O(\cosh (2r))$. Remembering that $\det M =-\eta^2$, we find again 
\begin{equation}
 \lim_{r\rightarrow \infty} 2\nu_-^2 = \frac{2 \det N}{(1+\det M)^2}.
\end{equation}
Moreover, combining this expression with the condition (\ref{cpmap}) of complete positivity, one finds that for such channel $\nu_-$ is always larger than one, i.e., $E_N[\rho_{\mathrm{out}}]=0$, so that the output state $\rho_{\mathrm{out}}$ can never be entangled. These channels are called entanglement breaking channels. Finally, the last class of channels, characterized by $\det M =0$, can easily be proven to be entanglement breaking using similar arguments, which completes the proof. 
$\blacksquare$

We are now in a position to prove the main result of the paper. 

\textbf{Theorem:} Given a Gaussian channel $T$ with matrices $M$ and $N$, there exists no GECC that transforms $T$ into a Gaussian channel $T_\mathrm{GC}$ with matrices $M_\mathrm{GC}$ and $N_\mathrm{GC}$ having a lower entanglement degradation, i.e., such that 
\begin{equation}\label{theo}
\frac{\det N_\mathrm{GC}}{(1+\det M_\mathrm{GC})^2}< \min\left\{\frac{\det N}{(1+\det M)^2},1\right\}\, . 
\end{equation}

\textbf{Proof.} Our proof works by contradiction. Suppose that there exists a GECC as in Fig. \ref{fig:gec} whose overall effect is to transform $T$ into $T_{GC}$, and such that the condition (\ref{theo}) is satisfied. By Lemma 1, there exists a one-way Gaussian protocol as in Fig. \ref{fig:ed} which transforms $n$ copies of the state $\chi$ into the state 
$\rho_{d}=\lim_{r\rightarrow \infty} \identity\otimes T_\mathrm{GC}(\proj{\phi_{r}})$.
Lemma 2 combined with condition (\ref{theo}) shows that
$ E_N[\rho_{d}]>E_N[\chi]$; hence, the resulting one-way protocol is a true entanglement distillation protocol based on Gaussian operations only. This is in clear contradiction with the impossibility to distill entanglement of a Gaussian state with Gaussian operations \cite{note}. We conclude that such a GECC does not exist. $\blacksquare$ 

We now illustrate this no-go theorem by applying the criterion (\ref{theo}) to some well-known Gaussian channels.

\textit{Attenuation channel.} This channel $T_\eta$ is characterized by $M=\eta\identity$ and $N=|1-\eta^2|\identity$, with $\eta<1$. It is the prototype channel for optical communication through a lossy fiber, and can be modeled by a beam splitter of transmittance $\eta$. Its entanglement degradation  
\begin{equation}\label{Plossy}
 D[T_\eta]=\frac{(1-\eta^2)^2}{(1+\eta^2)^2}<1
\end{equation}
is a decreasing function of $\eta$. Hence, by (\ref{theo}), it is impossible to find a GECC that turns $T_\eta$ into another attenuation channel with less losses. As is well known, one can nevertheless reduce the attenuation ($\eta<\eta_\mathrm{GC}<1$)
but at the expense of an increasing noise $N_\mathrm{GC}$. A bound on the minimum achievable noise is given by (\ref{theo}).

\textit{Amplification channel.} It is similar to the attenuation channel, but with $\eta>1$. Thus, Eq. (\ref{Plossy}) holds, but now $D$ is an increasing function of $\eta$, so that it is impossible to make $\eta_\mathrm{GC}<\eta$. Again, one can reduce the amplification at the expense of an increased noise, e.g., by concatenating the amplification channel with an attenuation channel.

\textit{Classical noise channel.} This channel $T_N$ adds Gaussian classical noise to the input state, i.e., $M=\identity$ and $N>0$. Its entanglement degradation is
\begin{equation}
 D[T_N]=\min\left\{\frac{\det N}{4},1\right\},
\end{equation}
so that our theorem implies that it is impossible to reduce the noise when $\det N\leq4$, and that it is impossible to reduce the noise under $4$ when $\det N > 4$. Note that this limit of 4 can always be reached as the number of available ancillas goes to infinity. Alice simply needs to optimally measure the input state and send to Bob an infinite number of states centered on her measurement result. Bob measures the received states and prepares a state centered on the average value of his measurements. This measure-and-prepare strategy yields $\det N = 4$.

To summarize our results, we have introduced an intrinsic property of single-mode Gaussian channels called the entanglement degradation, and proven that it cannot be reduced by Gaussian encoding and decoding operations only. Such a no-go theorem for the correction of Gaussian errors with Gaussian operations nicely complements the well-known impossibility to distill the entanglement of Gaussian states with Gaussian local operations. 
In this paper we focused on deterministic Gaussian CP maps since these describe the most
common practical communication channels.  However, the no-go theorem can be straightforwardly
extended to probabilistic trace-decreasing Gaussian CP maps isomorphic to generic two mode Gaussian states $\chi$ 
\cite{distil2,distil3}. One only needs to define $D[T]=\min(1,\nu_{-}^2)$, where $\nu_{-}$ is the lower symplectic eigenvalue of covariance matrix of partially transposed $\chi$. 

sInterestingly, the entanglement degradation can be related to another important intrinsic properties of channels, namely the quantum capacity $Q$. In particular, one can show that the quantum capacity of a single-mode Gaussian channel $T$ is always upper bounded by the function of $D$ 
\begin{equation}
 Q[T]\leq -\frac{1}{2}\log D[T]\, .
\end{equation}
This result follows from \cite{WH}, where a computable upper bound on the quantum capacity was introduced. This capacity-like quantity $Q_{\theta}$ can be defined as the maximal entanglement, as measured by the logarithmic negativity, of states transmitted through the channel $T$, i.e., $Q_{\theta}[T]=-\frac{1}{2}\log D[T]$ for single-mode Gaussian channels. 
A natural and promising extension of our paper would therefore be to investigate whether a Gaussian no-go theorem 
also holds for the quantum capacity of Gaussian channels.

We acknowledge financial support of the EU under the FET-Open project COMPAS (212008). J.N. acknowledges support from the Belgian FRIA foundation, and J.F. acknowledges support from MSMT (grants No. LC06007 and No. MSM6198959213).


\begin{thebibliography}{99}
\bibitem{note}
Note that in \cite{distil3}, the impossibility is proven using $V(\gamma_{AB})$ as the entanglement measure, where $V(\gamma_{AB})$ is defined as the largest value $p\leq 1$ such that $\gamma_{AB}\geq p(\gamma_A \oplus \gamma_B)$ for some $\gamma_A$ and $\gamma_B$. However, it is shown in the appendix of \cite{distil3} that for a $1 \times N$ Gaussian state $\rho_{AB}$, $V(\gamma_{AB})$ is exactly equal to the smallest symplectic eigenvalue of the covariance matrix of the partially transposed state $\rho_{AB}^{T_A}$. It follows that $E_N(\gamma_{AB})$ is a monotonic function of $V(\gamma_{AB})$, so that the result of \cite{distil3} is equivalent to the impossibility of increasing the logarithmic negativity of $1 \times N$ Gaussian states with Gaussian operations only.
\end{thebibliography}
\end{document}